\documentclass[
aps,
prr,
reprint,
superscriptaddress,
amsmath,
amssymb
]{revtex4-2}
\usepackage{graphicx}
\usepackage{hyperref}
\usepackage{amsthm}

\newcommand{\dbar}{{\mathchar'26\mkern-11mu}\mathrm{d}}
\newcommand{\Sr}{Sr$_2$RuO$_4$}

\begin{document}

\title{Entropy mapping under uniaxial pressure utilizing the elastocaloric effect}

\author{Zhenhai Hu}
\affiliation{Max Planck Institute for Chemical Physics of Solids, N{\"o}thnitzer Str.\ 40, 01187 Dresden, Germany}
\affiliation{Scottish Universities Physics Alliance, School of Physics and Astronomy, University of St Andrews, St Andrews KY16 9SS, UK}

\author{You-Sheng Li}
\affiliation{Department of Physics, National Taiwan University, Taipei City 106319, Taiwan (R.O.C.)}
\affiliation{Max Planck Institute for Chemical Physics of Solids, N{\"o}thnitzer Str.\ 40, 01187 Dresden, Germany}

\author{Aleksei V. Frolov}
\affiliation{Max Planck Institute for Chemical Physics of Solids, N{\"o}thnitzer Str.\ 40, 01187 Dresden, Germany}

\author{Fabian Jerzembeck}
\affiliation{Max Planck Institute for Chemical Physics of Solids, N{\"o}thnitzer Str.\ 40, 01187 Dresden, Germany}

\author{Manuel Brando}
\affiliation{Max Planck Institute for Chemical Physics of Solids, N{\"o}thnitzer Str.\ 40, 01187 Dresden, Germany}

\author{Naoki Kikugawa}
\affiliation{National Institute for Materials Science, Tsukuba 305-0003, Japan}

\author{Dmitry A. Sokolov}
\affiliation{Max Planck Institute for Chemical Physics of Solids, N{\"o}thnitzer Str.\ 40, 01187 Dresden, Germany}

\author{Hilary M. L. Noad}
\affiliation{Max Planck Institute for Chemical Physics of Solids, N{\"o}thnitzer Str.\ 40, 01187 Dresden, Germany}

\author{Andrew P. Mackenzie}
\affiliation{Max Planck Institute for Chemical Physics of Solids, N{\"o}thnitzer Str.\ 40, 01187 Dresden, Germany}
\affiliation{Scottish Universities Physics Alliance, School of Physics and Astronomy, University of St Andrews, St Andrews KY16 9SS, UK}

\author{Michael Nicklas}
\affiliation{Max Planck Institute for Chemical Physics of Solids, N{\"o}thnitzer Str.\ 40, 01187 Dresden, Germany}

\author{Andreas W. Rost}
\email{a.rost@st-andrews.ac.uk}
\affiliation{Scottish Universities Physics Alliance, School of Physics and Astronomy, University of St Andrews, St Andrews KY16 9SS, UK}
\affiliation{Max Planck Institute for Chemical Physics of Solids, N{\"o}thnitzer Str.\ 40, 01187 Dresden, Germany}

\date{\today}

\begin{abstract}
Uniaxial pressure is a powerful tuning parameter for quantum materials, but conventional thermodynamic probes such as specific heat are difficult to realize in the constrained geometries of strain apparatus. We develop a quantitative analysis framework for a.c.\ elastocaloric effect measurements that enable the reconstruction of the absolute entropy and hence specific heat across complex phase diagrams. The absolute accuracy is achieved by combining measurements in the strong coupling regime at low frequencies with high signal-to-noise measurements in the quasi-adiabatic regime at high frequencies. Applying the approach to the correlated superconductor \Sr, we obtain an absolute entropy map across the phase diagram including across phase transitions deep into the superconducting state. We demonstrate that from such data one can derive the absolute specific heat which is currently not possible through other approaches. This data reinforces the finding that the quenching of entropy within the superconductor \Sr ~is strongest at the critical strain consistent with the superconducting gap being maximized at the Van Hove singularity (VHs). Furthermore, we demonstrate that, although $\Delta c /(\gamma T) $ does increase at the VH strain, this increase is much weaker than previously inferred from more indirect caloric experiments.
\end{abstract}

\maketitle

\section{Introduction}

Uniaxial strain has become an important tool for studying the pressure phase diagrams of quantum materials~\cite{Hicks2025} with examples including iron pnictide~\cite{Chu2010,Chu2012,Worasaran2021} and cuprate superconductors~\cite{Kim2018, Vinograd2024}, charge density wave compounds~\cite{Nicholson2021,straquadine2022,Welp1992} and semimetals~\cite{Jo2019}. Notably the development of piezo-driven rigs~\cite{HicksRSI} has enabled rapid progress in this field. However, the restrictive sample conditions in such experiments constrain the implementation of absolute thermodynamic measurements. In particular, the study of entropy changes across the phase diagram using specific heat as a key experiment is extremely challenging. Due to the thermal clamping of the sample, the quantitative analysis of standard techniques such as quasi-adiabatic a.c.\ or relaxation time based methods has large systematic uncertainties requiring extensive modeling ~\cite{nature,Li2018}. As a result the a.c.\ elastocaloric effect, which reflects entropy changes as a function of strain rather than temperature, has been explored in several works as a means to establish the thermodynamics of phase diagrams~\cite{nature,IkedaRSI1,IkedaRSI2}.
The focus has so far been on exploiting the high-frequency quasi-adiabatic regime in which the signal-to-noise is maximized. However, the extraction of absolute elastocaloric effect data in this frequency regime faces the same systematic challenges as specific heat measurements. In addition, the quasi-adiabatic signal reflects changes of the entropy both with strain and temperature that are challenging to disentangle across phase transitions~\cite{nature}. Combined these constraints can result in a significant systematic uncertainty across a phase diagram~\cite{IkedaRSI2}.

In this paper we present an approach to overcome these challenges by combining a.c.\ measurements in two different frequency regimes. This enables us to reconstruct an accurate and precise map of the elastocaloric effect of such quality that a calculation of the absolute entropy and specific heat changes across phase transitions is possible.

\begin{figure}[tb!]
\centering
\includegraphics[width=\linewidth]{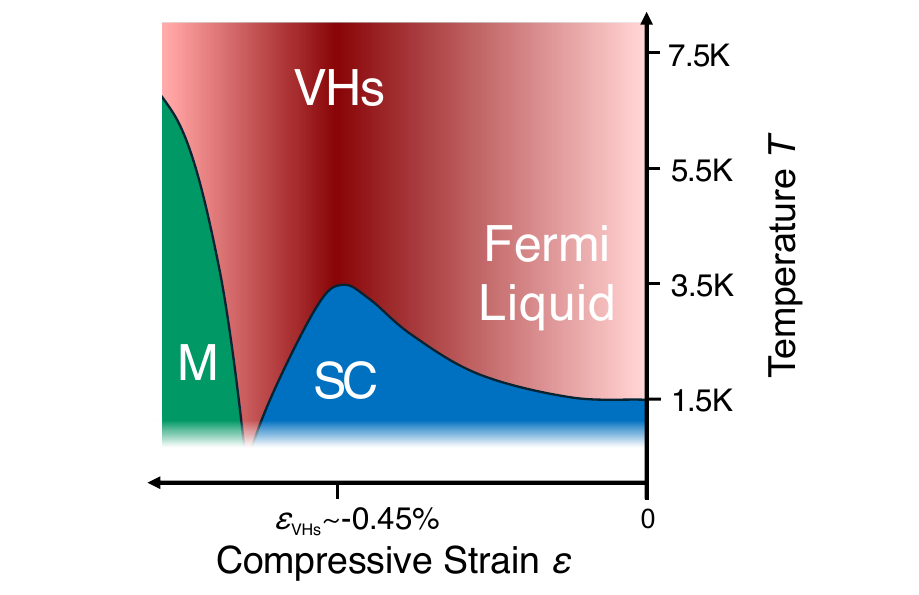}
\caption{Schematic temperature--strain phase diagram of Sr$_2$RuO$_4$ as a function of compressive strain $\varepsilon$, along the $\langle 100 \rangle$ crystal axis.
The superconducting (SC) dome is enhanced in the vicinity of the strain-tuned Van Hove
singularity (VHs) at $\varepsilon_{\mathrm{VHs}}~$\cite{nature}, separating the low-temperature superconducting
phase from the normal-state which at low strains is well described by standard Fermi liquid theory.
At larger compressive strain, a magnetically ordered (M) phase emerges~\cite{Grinenko2021}.
At $\varepsilon_{\mathrm{VHs}}$, the increased darker shading denotes an enhanced density of states.}
\label{fig:PD}
\end{figure}

\begin{figure*}[t!]
\centering
\includegraphics[width=\linewidth]{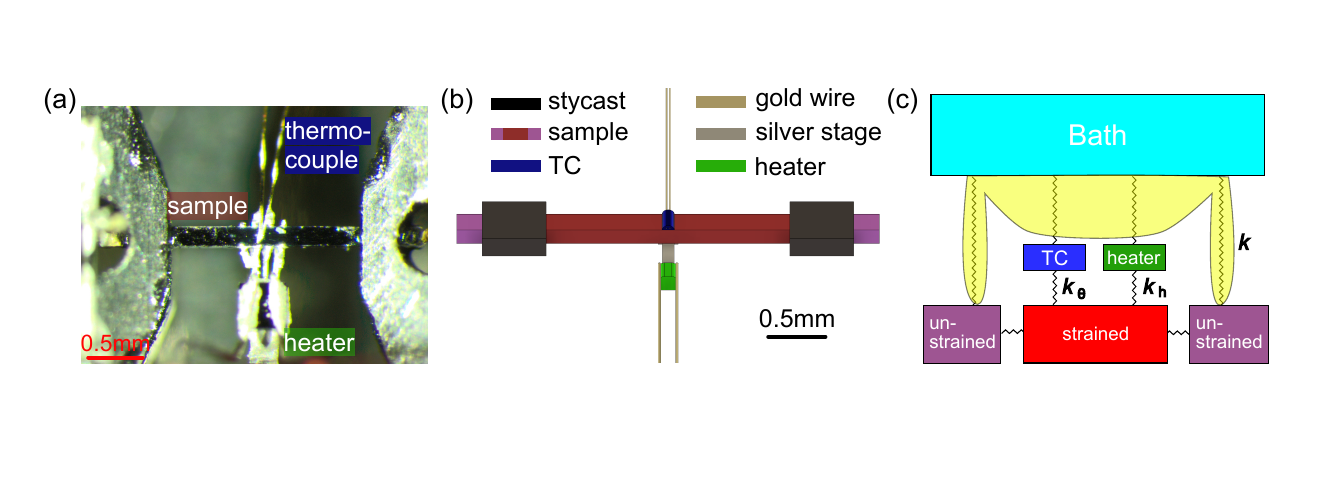}
\caption{Experimental setup and basic thermal model for heat-capacity and elastocaloric effect measurements.
(a) Photograph of the experimental setup with the principal components annotated. An Au/AuFe thermocouple and a resistive heater are thermally coupled to the sample using silver epoxy, while the sample is mechanically anchored to titanium plates with epoxy resin.
(b) Schematic diagram of the setup, emphasizing the spatial arrangement and thermal connections between the sample, heater, thermocouple, and strain cell.
(c) Abstracted lumped-element thermal model for setup in (a) and (b). The thermocouple and heater are strongly coupled to the sample via thermal conductances \(k_{\theta}\) and \(k_{h}\), forming an effectively isothermal subsystem. This subsystem is weakly coupled to the thermal bath through an effective thermal conductance \(k\).}
\label{fig:fig1}
\end{figure*}

We apply this approach to the uniaxial pressure phase diagram of \Sr\,  schematically shown in Fig.~\ref{fig:PD}. This tetragonal material has been extensively studied under uniaxial strain along the $\langle100\rangle$ direction for which it is possible to tune an electronic Van Hove singularity (VHs) across the Fermi energy  at $\varepsilon_\mathrm{VHs}=-0.45\,\%$ ~\cite{Steppke2017,Sunko2019}. At zero strain, a superconducting phase transition is observed below 1.5\,K emerging out of a Fermi liquid normal state~\cite{Mackenzie2003}. For compressive strains the superconducting transition temperature increases to approximately 3.5\,K close to $\varepsilon_\mathrm{VHs}$ before sharply dropping.
At higher compressive strains a magnetic phase is stabilized~\cite{Grinenko2021,nature}. In a previous elastocaloric effect study we reconstructed the entropy $S$ above the superconducting phase transition and demonstrated that it peaks close to $\varepsilon_\mathrm{VHs}$ ~\cite{nature}, behavior consistent with an enhanced electronic density of states crossing the Fermi energy at $\varepsilon_\mathrm{VHs}$.
In this paper we develop a new analysis approach that enables us to extend the reconstruction of the entropy across phase transitions and to low temperatures well into the superconducting phase by overcoming systematic uncertainties in the previous analysis. Establishing the evolution of the absolute entropy in \Sr\, across this phase diagram will provide quantitative data against which to test the various competing superconducting order parameters in this compound~\cite{Palle2023}. More broadly, we show that the method we introduce yields absolute elastocaloric data with an overall uncertainty of the order of $30\%$. This uncertainty is dominated by the difficulty to accurately determine the strained sample volume.

\section{Methods}

\subsection{Experimental Setup}

The experiments reported here are based on the setup reported in~\cite{Li2021, nature}, a photograph of which is shown in Fig.~\ref{fig:fig1}(a). A schematic highlighting the key components is presented in Fig.~\ref{fig:fig1}(b) together with the resulting thermodynamic model in Fig.~\ref{fig:fig1}(c).

The strain cell allows for the simultaneous application of a d.c.\ and a.c.\ strain~\cite{nature}. The strain modulation amplitude was adjusted between \(2.9\times10^{-6}\) and \(1.6\times10^{-5}\), depending on the measurement conditions, in order to maintain the temperature oscillation within an optimal and measurable response range. The sample has dimensions of approximately $W= 200~\mu$m, $H= 150~\mu$m, $L= 3.4$~mm and is mounted on the clamps of the strain cell with Stycast 2850FT epoxy resin approximately $20~\mu$m thick. The strained portion (between two clamping plates) is approximately 2~ mm in length. The sample temperature is measured with an Au/AuFe thermocouple attached with high temperature cured Dupont 6838 silver epoxy. On the opposite side a heater is attached similarly to the sample via four silver wires. A more detailed description is given in Refs.~\cite{Li2018,Li2020,nature}.

\subsection{Thermodynamic Response}

In Fig.~\ref{fig:fig1}(c) we present the lumped-element model of the experiment representing all key elements and relevant thermal links (lumped elements as relevant physical properties of individual elements / thermal links are represented by single scalar values). This approach is justified if the thermal length~\cite{ACmodel} of the elements is longer than the physical dimensions and temperature gradients across thermal barriers are linear~\cite{IkedaRSI2}. These conditions are fulfilled in the low-frequency strong-coupling regime that is the focus of this paper but in principle do have to be considered in a quantitative analysis of the quasi-adiabatic regime at high frequencies.

Although a strong simplification, the finite frequency response of the model can already be illustrated in a basic version assuming negligible heat capacities of the thermocouple and heater as well as these being perfectly coupled to the sample.
These approximations leave the sample's total entropy $S_\mathrm{tot}$, which is the sum of the entropy of the strained part ($S_\mathrm{strain}$) and unstrained part ($S_\mathrm{unstrained}$), the thermal link $k$ of the sample to the bath, as well as the heat flow $Q_\mathrm{link}$ from the sample to the bath as the only relevant variables. As we show in Appendix~\ref{app:model} the resulting model can be solved analytically. In Appendix~\ref{app:model} we also discuss the relationship of the a.c.\ elastocaloric analysis presented here to the more widely known a.c.\ specific heat analysis, which requires consideration of the heat $Q_\mathrm{heater}$ deposited by the heater.

In the model described above an applied a.c.\ elastocaloric oscillation $\tilde\varepsilon$ results in temperature oscillations of magnitude  $\left|\tilde{T}_s\right|$ as a function of frequency $\omega$ given by
\begin{equation}
\label{solutionmag}
\left|\tilde{T}_s\right|=
\frac{\omega}{\sqrt{1+\omega^2\left(C_\mathrm{tot}/k\right)^2}}\times\frac{T_0}{k}\left|\left(\frac{\partial S_\mathrm{strain}}{\partial \varepsilon}\right)_{T}\tilde{\varepsilon}\right| .
\end{equation}

\begin{figure}[b]
\centering
\includegraphics[width=\linewidth]{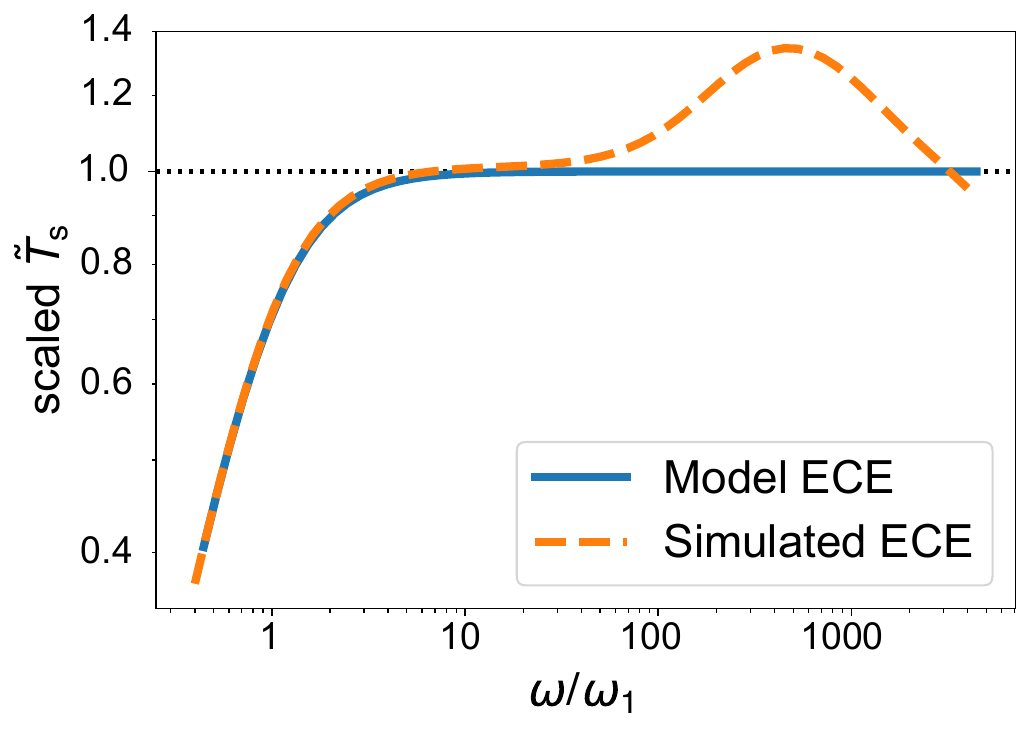}
\caption{Normalized $\tilde{T}_s$ scaled by quasi-adiabatic response.
Scaled $\tilde{T}_s=$1 corresponds to the idealized limit of a perfectly adiabatic and isothermal response of the whole sample.
The blue curve shows the analytical result given by Eq.~\ref{solutionmag},
while the orange curve corresponds to a numerical simulation incorporating
a more realistic experimental thermal model.
Both are plotted as functions of the dimensionless drive frequency
$\omega/\omega_1$, where $\omega_1$ is defined in text.}
\label{fig:fig3}
\end{figure}

\noindent Here $C_\mathrm{tot}$ refers to the total heat capacity which is the sum of the contributions from the strained and unstrained parts of the sample and $T_0$ is the average temperature. The behavior described by Eq.~\ref{solutionmag} has a characteristic time / frequency scale set by $\tau_{\rm 1}=1/\omega_{\rm 1}=\frac{C_\mathrm{tot}}{k}$. The functional form is shown in blue in Fig.~\ref{fig:fig3}.

The characteristic frequency separates the response into two well defined regimes. In the low-frequency regime $\omega\tau_{\rm 1}\ll1$ Eq.~\ref{solutionmag} simplifies to
\begin{equation}
\label{solutionmag-lf}
\left|\tilde{T}_s\right|=\omega\times\frac{T_0}{k}\left|\left(\frac{\partial S_\mathrm{strain}}{\partial\varepsilon}\right)_T\tilde{\varepsilon}\right|,
\end{equation}

\noindent whereas in the high-frequency regime $\omega\tau_1 \gg1$ it simplifies to

\begin{equation}
\label{solutionmag-hf}
\left|\tilde{T}_s\right|=
     \frac{T_0}{C_\mathrm{tot}}\times\left|\left(\frac{\partial S_\mathrm{strain}}{\partial \varepsilon}\right)_T\tilde{\varepsilon}\right| .
\end{equation}

The latter is often referred to as the quasi-adiabatic regime as the temperature oscillations are so fast that little to no heat is exchanged with the bath throughout the cycle. In the low frequency regime, on the other hand, the temperature deviation of the sample from the bath temperature are strongly suppressed as a result of which we refer to it as the strong coupling regime.

The elastocaloric signal size is maximized in the quasi-adiabatic regime which is why it is commonly chosen for measurements. However, the above analysis neglects several aspects of the experiments that in particular affect the quasi-adiabatic regime. We highlight this in the orange model curve in Fig.~\ref{fig:fig3}, which includes the effects of finite sample length and an imperfect coupling of the thermometer for realistic parameters of our setup. The thermal length of the sample becoming comparable to the geometric dimensions results at high frequencies in a decoupling of the unstrained parts from the strained parts. As a consequence of this decoupling less of the sample is being heated / cooled during the elastocaloric cycle, which in turn leads to an overshoot of the temperature oscillations \cite{IkedaRSI2}. In Fig.~\ref{fig:fig3} this increased signal occurs for frequencies above $\omega/\omega_1>10$. The finite thermal length also results in nonlinear temperature gradients within the sample. Furthermore, at high frequencies the thermocouple decouples from the sample due to the finite thermal link between sample and thermometer, resulting in a drop of the measured signal. This occurs for the orange model trace shown in Fig.~\ref{fig:fig3}  for frequencies of $\omega/\omega_1>300$.

Including these two effects already leads to rather large systematic uncertainties in the analysis of the quasi-adiabatic regime as $\omega_1$ and the other characteristic frequencies shift significantly across the measured phase diagram. Such shifts are particularly large at phase transitions with $\omega_1$ varying by as much as a factor of two. Hence generally there is no single frequency at which measurements would be in the quasi-adiabatic regime across the whole phase diagram. Consequently, the reconstruction of entropy and specific heat from quasi-adiabatic data, which relies on integration and differentiation of the measured signal, becomes unreliable. In addition, the measurement signal in the quasi-adiabatic regime depends on the {\it a priori} unknown heat capacity $C_{\rm tot}$ of the sample (see Eq.~\ref{solutionmag-hf}). This dependency significantly complicates the extraction of quantitative thermodynamic data from the experiment \cite{nature}.

The strong coupling regime, in contrast, is in general insensitive to these experimental details, as the measurement frequency $\omega$ is far below $\omega_1$ and the other characteristic frequencies above which the various decoupling mechanisms set in. As a consequence, this is the measurement regime that should be utilized in order to accurately extract the absolute magnitude of $\left(\partial S_\mathrm{strain}/\partial \varepsilon\right)_T$ from elastocaloric effect measurements using Eq.~\ref{solutionmag-lf}.
However, the signal-to-noise ratio in this regime is often insufficient for further analysis, as the signal is intrinsically at least a factor of three smaller than in the quasi-adiabatic regime.
In addition, the noise floor generally increases towards lower frequencies, such that the overall signal-to-noise ratio can be an order of magnitude lower compared to the quasi-adiabatic regime.

\begin{figure}
\centering
\includegraphics[width=\linewidth]{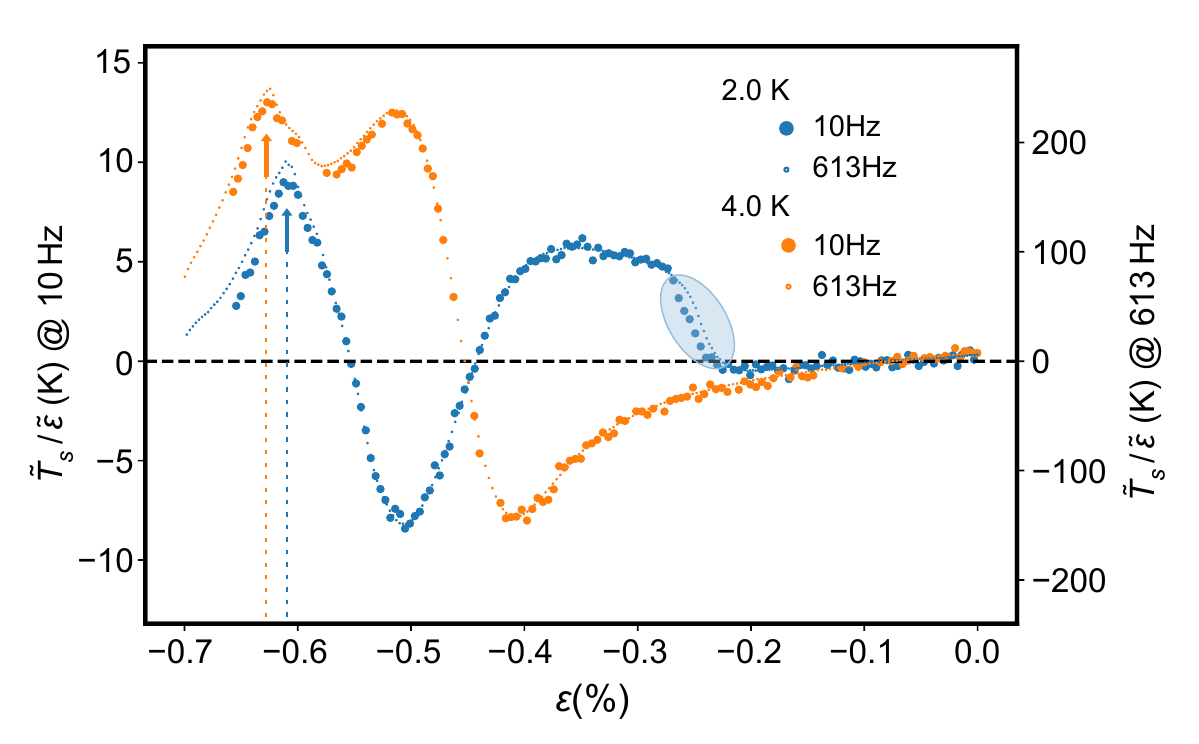}
\caption{Strain dependence of the elastocaloric signal in Sr$_2$RuO$_4$.
The elastocaloric effect $\tilde{T}_s/\tilde{\varepsilon}$ is shown as a function of
strain $\varepsilon$ at temperatures of 2.0\,K and 4.0\,K, measured at modulation
frequencies of 10\,Hz (left scale) and 613\,Hz (right scale).}
\label{fig:fig3b}
\end{figure}

In Fig.~\ref{fig:fig3b}, we illustrate this inherent trade-off between absolute accuracy (strong-coupling regime) and sensitivity (quasi-adiabatic regime). For our setup the characteristic frequency $\omega_1/2\pi$ is of order $100$\,Hz. Hence, we present the amplitude of the temperature oscillations as a function of d.c.\ strain both at temperatures of $T=$ 2\,K (blue) and $T=$ 4\,K (orange), deep within the strong-coupling regime ($\omega/2\pi= 10$\,Hz - large circles) and the quasi-adiabatic regime ($\omega/2\pi= 613$~Hz - small circles). The sign indicates if the temperature oscillations correspond to an increase (positive) or decrease (negative) in entropy with increasing strain $\varepsilon$. Data at $T=$ 4\,K (orange) are taken above the superconducting dome in the normal state. The measurement at $T=$ 2\,K (blue) crosses superconducting phase transitions at low and high compressive strains (e.g. shaded region) as well as the magnetic phase at $\varepsilon_{M}$ as indicated by arrows.

The signal-to-noise ratio between the two measurement frequencies varies by more than an order of magnitude (note the different $y$-scale for the measurements at different frequencies). In addition, there are significant systematic differences when crossing phase transitions as highlighted by the differences in the 2\,K trace at $\varepsilon\approx-0.25\%$ (blue shaded region). While apparently small, such differences substantially impact the reconstruction of the entropy and hence specific heat from such data.

While there are significant differences the overall close correlation between the strong-coupling and quasi-adiabatic is striking. This correlation is key to the subsequent analysis as it implies that the strong-coupling low frequency oscillations $T_{LF}$ and quasi-adiabatic high frequency oscillations $T_{HF}$ can be related by a smoothly varying mapping function of the form

\begin{equation}
\label{eq:mapping}
    T_{LF}(\varepsilon,T)=A(T)\times T_{HF}+B(\varepsilon,T),
\end{equation}

\noindent such that the parameters $A(T)$ and $B(\varepsilon,T)$ have a much weaker temperature and strain dependence than $T_{LF}$ or $T_{HF}$ themselves. The underlying physical reason is that while the measurement signals can change by more than an order of magnitude across the phase boundaries the change of the ratio between $T_{LF}$ and $T_{HF}$ is only of order unity. This weak variation of the ratio between the strong-coupling and quasi-adiabatic data sets allows for the creation of a $\left(\partial S/\partial\varepsilon\right)_T$ dataset with both high accuracy and high signal-to-noise through a combination of the information contained in both data sets. To achieve this, we implement the following algorithm at each temperature:

\begin{enumerate}
    \item Establish the primary linear proportionality by fitting $ T_{LF}(\varepsilon) = A\times T_{HF}(\varepsilon) $ with $A$ a fit parameter.
    \item Compute: $ B(\varepsilon) =  T_{LF}(\varepsilon) - A\times T_{HF}(\varepsilon)$.
    \item Suppress high-frequency noise in $B(\varepsilon)$
    by applying a Savitzky--Golay filter,
    yielding $B_{\mathrm{F}}(\varepsilon)$.
    \item Reconstruct:
    \begin{equation}
    \label{eq:mapping2}
        \hat{T}(\varepsilon ) = A\times T_{HF}(\varepsilon ) + B_{\text{F}}(\varepsilon ).
    \end{equation}
\end{enumerate}

The important difference between smoothing $B$ and adaptive smoothing of the low frequency data $T_{LF}$ is that $B$ is a function with significantly less variation as a function of strain. As we will demonstrate, the resulting dataset does allow for the calculation of $S(\varepsilon,T)$ with sufficient accuracy and sensitivity to enable the calculation of, e.g., the specific heat and resolve fine features in the vicinity of optimal $T_{\mathrm{c}}$.

\subsection{Calibration in Strong-Coupling Regime}

Inspection of Eq.~\ref{solutionmag-lf} reveals two further points to be addressed in order to extract the material-specific elastocaloric effect from a measurement in this regime. The first is independent knowledge of the thermal conductance to the bath $k$, which in our case is dominated by the epoxy-resin layer. This point can be addressed experimentally. A known heat $Q_{heater}$ can be delivered to the sample by the heater and the temperature difference with the bath $\Delta T$ measured by the thermocouple giving $k=Q/\Delta T$.

Secondly the amount of \Sr~contributing to the strained part of the sample has to be determined accurately. In general it is possible to simply estimate the strained volume based on the sample geometry.
Given the intrinsic uncertainties due to the clamping, strain inhomogeneity and overall sample dimensions we estimate that this has in general a systematic error of at least 10\%.

As discussed in Ref.~\cite{nature} in the case of \Sr\ a complementary method utilizing the known elastic properties in the Fermi liquid regime at low strains and above the superconducting transition can be exploited. Following the nomenclature of our previous work \cite{nature}, in this part of the phase diagram, the volumetric specific heat (i.e., heat capacity divided by reference volume at zero strain) is described by

\begin{equation}
C_V(\varepsilon, T) = \gamma \left( 1 + \varepsilon \gamma_1 / \gamma + \varepsilon^2 \gamma_2 / \gamma \right) T + \beta T^3 ,
\end{equation}

\noindent with $\gamma$ being the Sommerfeld coefficient, $\gamma_1$ and $\gamma_2$ describing variations of the Sommerfeld coefficient linear and quadratic in strain, and $\beta T^3$ the usual phonon contribution \cite{Mackenzie2003,nature}. Integration as a function of temperature yields the entropy
\begin{equation}
\label{eq:entropy3}
\begin{array}{rcl}
S_V(\varepsilon, T) = \left( \gamma + \gamma_1 \varepsilon + \gamma_2 \varepsilon^2 \right) T + \frac{1}{3} \beta T^3,\\
\end{array}
\end{equation}

Furthermore given that stress $\sigma$ and strain $\varepsilon$ are related by means of the compliance matrix $\mathbf{s}$ \cite{entropy_compliance}  through
\begin{equation}
\varepsilon = \mathbf{s} \sigma,
\end{equation}
one can show via Maxwell relationships \cite{nature} that
\begin{equation}
\label{eq:entropy2}
\left( \frac{\partial^2 S_V}{\partial \varepsilon^2}  \right)_{T}
= -\left( \frac{\partial}{\partial T} s_{11}^{-1} \right)_{\varepsilon}
= 2\gamma_2T \, ,
\end{equation}

\noindent with $s_{11}$ being the 11 entry of the compliance tensor.
Hence

\begin{equation}
\label{eq:entropycalc}
\left(\frac{\partial S_V}{\partial\varepsilon}\right)_T=  \gamma_1 T - \varepsilon\left( \frac{\partial}{\partial T} s_{11}^{-1} \right)_{\varepsilon, \sigma_y, \sigma_z},
\end{equation}
which in combination with Eq.~\ref{solutionmag-lf} results in the amplitude of a.c.\ elastocaloric temperature oscillations in the strong-coupling regime being described by

\begin{equation}
\label{eq:fitfunctioncalib}
\left|\tilde{T}_s\right|=\omega\frac{T}{k}\frac{n}{\rho_M}\left( \gamma_1 T - \varepsilon\left( \frac{\partial}{\partial T} s_{11}^{-1} \right)_{\varepsilon
}\right) \left|\tilde{\varepsilon}\right|,
\end{equation}
where we used the molar density $\rho_M$ to convert volume to molar sample amount $n$. This final equation completely describes the response in the low strain Fermi liquid regime with the two unknowns $n$ and $\gamma_1$. Hence a fit of Eq.~\ref{eq:fitfunctioncalib} to the data in the Fermi-liquid regime of \Sr~ between, e.g., temperatures of 2 and 4~K and strains between $-0.1$ and 0\% allows for a determination of the amount of strained sample complementary to that based on strained sample volume.

\begin{figure*}[!t]
\centering
\includegraphics[width=\linewidth]{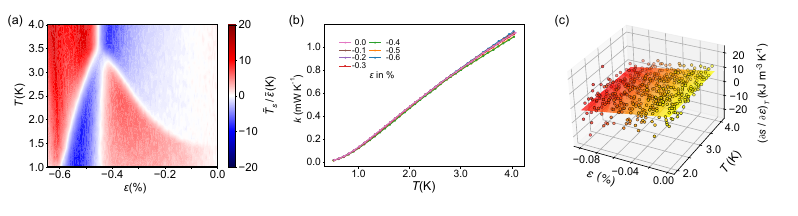}
\caption{Calibration of the low-frequency elastocaloric effect.
(a) Elastocaloric signal $\tilde T_s/\tilde\varepsilon$
as a function of temperature and strain.
(b) Temperature dependence of the reconstructed thermal conductance $k$ to the bath
at selected fixed strains.
(c) $(\partial s/\partial \varepsilon)_T$ in the Fermi liquid regime prior to calibration, together with the fitted calibration plane.}
\label{fig:calibration-lf}
\end{figure*}

\subsection{Reconstructing Entropy and Specific Heat}

The molar entropy $s$ \cite{entropy_compliance}  as a function of strain $\varepsilon$ and temperature $T$ can be formally expressed as the sum of the entropy at zero strain and its isothermal change as a function of $\varepsilon$. With the latter being the integral of $\left(\partial s/\partial\varepsilon\right)_T$ one therefore has
\begin{equation}
    \label{eq:totent}
    s\left(\varepsilon,T\right)= s \left(\varepsilon=0,T\right)+\int_{0}^\varepsilon\left(\frac{\partial s}{\partial \varepsilon'}\right)_T\mathrm{d}\varepsilon'\, .
\end{equation}
In the case of \Sr\, we furthermore know that at zero strain the material behaves as a perfect Fermi liquid above the superconducting transition. Hence knowledge of the molar specific heat $c$ at e.g.\ 4\,K fully determines the entropy at that temperature. Measurement of $c(\varepsilon=0)$ across the full temperature regime then allows for the calculation of zero-strain entropy at any relevant temperature as
\begin{equation}
\label{eq:entropyzero}
s\left(\varepsilon=0,T\right)=\gamma \times {\rm 4\,K}+\frac{1}{3}\beta\times ({\rm 4\,K})^3+\int_{\rm 4\,K}^T\frac{c}{T'}{\rm d}T' \, .
\end{equation}
The first two terms give the zero strain entropy at 4\,K resulting from the Fermi liquid behavior ($\gamma$-term) as well as a Debye phonon contribution ($\beta$-term)~\cite{Mackenzie1998}. The last term gives the entropy change relative to 4\,K based on the experimentally determined specific heat. This specific heat was determined in an independent dilution fridge based measurement on a sample of the same batch.

Equations~\ref{eq:totent} and \ref{eq:entropyzero} allow for the calculation of the absolute molar entropy across our whole measurement range. They involve a constant temperature integral of our data. As the data was taken along isotherms no interpolation of the dataset as a function of temperature is required.

\section{Results}
 In this section we present our results obtained on \Sr. We will first discuss the calculation of $\left(\partial s/\partial\varepsilon\right)_T$ from raw temperature oscillations at low and high frequencies, followed by the reconstruction of the entropy across the phase diagram and finally the resulting evolution of the specific heat.

\subsection{Calibration Elastocaloric Effect in Strong-Coupling Regime}

In Fig.\,~\ref{fig:calibration-lf}(a) we show a color-map of the magnitude of the temperature oscillations divided by the applied a.c.\ strain at a measurement frequency of 10\,Hz. The data were measured during isothermal sweeps of a d.c.\ `bias' strain at temperature intervals of 0.1\,K \footnote{Raw data was interpolated linearly and evaluated on a strain grid denser than the experimental data such that the figure gives a true representation of the noise levels.}.  At this frequency the setup is deep in the strong-coupling regime, i.e.\ $\omega\tau_{1}$ is well below 1 in Fig.~\ref{fig:fig3}.

In order to use Eq.~\ref{solutionmag-lf} we first determined the thermal link to the bath $k$. A basic estimate reveals the insulating epoxy resin layers to have a thermal conductance more than one order of magnitude lower than the unstrained \Sr~sample in the relevant temperature regime. Hence the epoxy resin completely dominates $k$ and provides the primary thermal resistance to the bath. As a result very little variation of $k$ with strain $\varepsilon$ is expected. This is borne out by the results of our temperature-dependent measurements presented for a select number of strains in Fig.~\ref{fig:calibration-lf}(b). Given the extremely weak strain dependence, the data could be well represented by a strain-independent curve for all strains. In the overall analysis we nevertheless incorporated this weak strain dependence.

Next we determine the amount of strained sample based on the Fermi liquid regime of the phase diagram. In Fig.~\ref{fig:calibration-lf}(c) we show a two-dimensional fit of Eq.~\ref{eq:fitfunctioncalib} to the data with the two unknown fit parameters being the molar amount of strained sample $n$ and $\gamma_1$. The fit results in $n=0.76\pm0.1$\,$\mu$mol and $\gamma_1=(6.1\pm0.5)\times10^3$\,Jm$^{-3}$K$^{-1}$. A geometric estimate based on the sample mounting would give $n=1.04$\,$\mu$mol, which can be considered a good agreement given the uncertainties of the optical distance measurements of the strained region of the mounted sample and strain inhomogeneities at the boundaries. In a more general context this confirms that even in samples without a Fermi liquid regime measurements in the strong coupling regime can give quantitative absolute results within 30~\% accuracy with the main uncertainty arising from the strained sample volume.

\begin{figure}[!h]
\centering
\includegraphics[width=\linewidth]{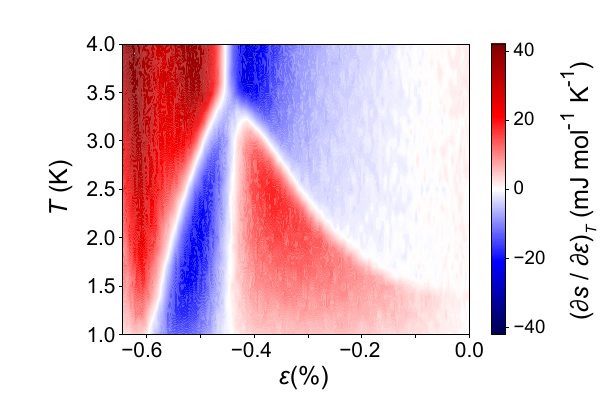}
\caption{Calibrated strain derivative of the entropy.
Color map of $\left(\partial s/\partial \varepsilon\right)_T$ as a function of temperature and strain.}
\label{fig:calibration-lf-2}
\end{figure}

Having thus determined all relevant factors to transform our temperature data to elastocaloric effect data we present in Fig.~\ref{fig:calibration-lf-2} the first key result of the paper which is $(\partial s / \partial \varepsilon)_T$ as function of strain and temperature across the phase diagram determined from strong-coupling regime measurements. While accurate in absolute value, we found that the signal-to-noise ratio is insufficient for the calculation, of e.g., the specific heat, which requires derivatives along the temperature axis.

\begin{figure*}
\centering
\includegraphics[width=\linewidth]{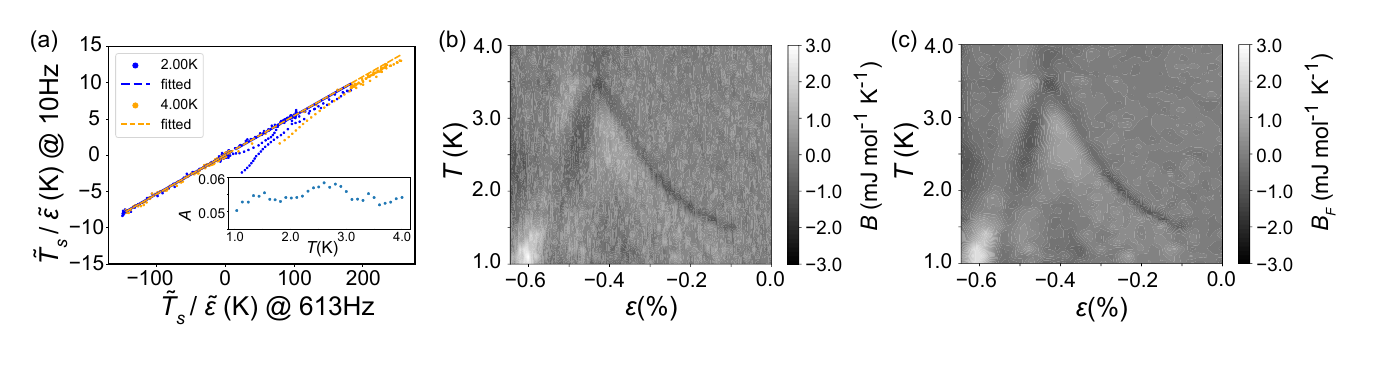}
\caption{Linear calibration procedure for low-frequency elastocaloric data.
(a) Linear correlation between the elastocaloric signals measured at high frequency (613~Hz)
and low frequency (10~Hz), shown for $T=2.0$~K and $4.0$~K; dashed lines indicate linear fits.
(b) Residual map, $B$, as a function of temperature and strain.
(c) Savitzky-Golay filtered residual map $B_F$, from which the final smoothing is obtained.}
\label{fig:fig5}
\end{figure*}

\subsection{Mapping of Quasi-Adiabatic Data}

To utilize the much higher signal-to-noise ratio of quasi-adiabatic measurements we took an equivalent dataset at a frequency of 613\, Hz. We decompose the relationship between the low frequency oscillations $T_{LF}$ and quasi-adiabatic high-frequency data $T_{HF}$ via Eq.~\ref{eq:mapping}. To illustrate how well the relationship between the datasets at different frequencies is described by a linear factor we re-plot the data from Fig.~\ref{fig:fig3b} in Fig.~\ref{fig:fig5}(a) where we show $T_{LF}$ as a function of $T_{HF}$. We fit Eq.~\ref{eq:mapping} at each temperature to determine $A(T)$ shown in the inset. In Fig.~\ref{fig:fig5}(b) we present the residual $B\left(\varepsilon,T\right)$ from these fits together with a Savitzky-Golay filtered surface $B_F\left(\varepsilon,T\right)$ in Fig.~\ref{fig:fig5}(c) on the same color scale. While this filtering removes most of the noise in the mapping function it is shape preserving in so far as the residual $B-B_F$ is featureless.

\begin{figure}[b!]
\centering
\includegraphics[width=\linewidth]{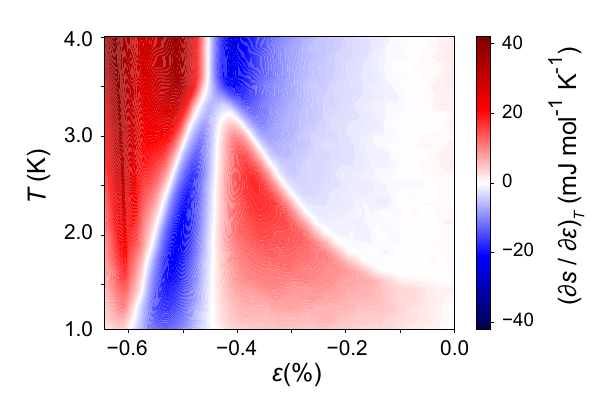}
\caption{Strain derivative of the entropy, $\left(\partial s/\partial \varepsilon\right)_T$,
as a function of temperature and strain.}
\label{fig:mappedfig6}
\end{figure}

One can now use Eq.~\ref{eq:mapping2} to map the quasi-adiabatic high-frequency data onto the strong coupling, low frequency data. Application to all our data results in a $\hat{T}(\varepsilon, T )$ that combines the high accuracy of the strong-coupling measurement at 10 Hz with the high signal-to-noise of our data set at 613 Hz. It is now possible using Eq.~\ref{solutionmag-lf} to calculate  $\left(\partial s/\partial\varepsilon\right)_T$ based on $\hat{T} $, which is presented in Fig.~\ref{fig:mappedfig6} as the second key result of this paper. The significantly higher signal-to-noise ratio compared to Fig.~\ref{fig:calibration-lf-2} is readily visible.

\subsection{Entropy and Specific Heat Reconstruction}

\begin{figure}[b!]
\centering
\includegraphics[width=\linewidth]{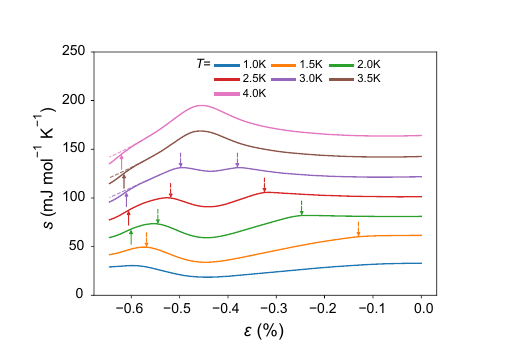}
\caption{Strain dependence of the entropy $s(\varepsilon)$ at selected temperatures.
Downward-pointing arrows mark the superconducting transitions and upward-pointing arrows mark the antiferromagnetic transitions. See text for details.}
\label{fig:ST}
\end{figure}

Next we calculate the absolute entropy across the phase diagram by numerically evaluating Eq.~\ref{eq:totent} using the data presented in Fig.~\ref{fig:mappedfig6}. In Fig.~\ref{fig:ST} we show the resulting molar entropy $s$ as a function of strain for a select number of temperatures.

Above 3.5\,K the entropy has a maximum close to $\varepsilon_\mathrm{VHs}\approx -0.45\%$ as qualitatively and quantitatively consistent with the analysis for the same regime in Ref.~\cite{nature}. The specific heat $c_{\varepsilon}$ across the phase diagram can be readily obtained by a numerical derivative of the reconstructed entropy with respect to temperature at constant strain as presented for select strains in Fig.~\ref{fig:HC}.

\section{Discussion}
The discussion here follows the data as presented in the result sections, first focusing on the elastocaloric effect data itself and then on the reconstructed entropy and the derived specific heat.

\subsection{Elastocaloric Effect}

To highlight the high precision of the reconstructed $\left(\partial s/\partial\varepsilon\right)_T$ data and agreement with theoretical analysis of the tight-binding model presented in Ref.~\cite{nature}, we focus on the region around the superconducting transition at optimal strain. In particular we are interested in whether the strain at which $T_{\mathrm{c}}$ is maximized coincides with the strain at which entropy is maximized.

In the color-map in Fig.~\ref{fig:mappedfig6} the color-bar is chosen such that white corresponds to $(\partial s/\partial\varepsilon )_T=0$. The maximum in entropy for temperatures above 3.5\,K is therefore easily identifiable by the vertical white line above 3.5\,K.

Furthermore, at optimal $T_{\mathrm{c}}$ it has to hold that $\mathrm{d} T_{\mathrm{c}} / \mathrm{d}\varepsilon$ is zero. Given the Ehrenfest relationship for second order phase transitions is

\begin{equation}
\label{eq:ehrenfest}
        \Delta \left( \frac{\partial s}{\partial\varepsilon} \right)_T = \frac{\Delta c_\varepsilon}{T_{\mathrm{c}}} \cdot \frac{\mathrm{d}{T}_{\mathrm{c}}}{\mathrm{d}\varepsilon}
\end{equation}

\noindent the vanishing of the derivative $\mathrm{d}{T}_{\mathrm{C}} / \mathrm{d}\varepsilon$ implies that $\Delta(\partial s/\partial\varepsilon )_T$---the difference between the elastocaloric effect in the normal and superconducting state---is zero at maximum $T_{\mathrm{c}}$.

If the maxima in $T_{\mathrm{c}}$ and in entropy in the normal state were to coincide at the same strain, then the Ehrenfest equation above would imply that the vertical white line continues into the superconducting state, as
$\Delta(\partial s/\partial\varepsilon)_T$ would have to be zero. This behavior is clearly not observed in the data. Indeed the maximum in $T_{\mathrm{c}}$ determined by Eq.~\ref{eq:ehrenfest} occurs at a compressive strain that is approximately 0.01\,\% smaller than the one at which the maximum in entropy occurs in the data. The same behavior is observed in the theoretical model discussed in Ref.~\cite{nature} where it is ascribable to the strong anisotropy of the Fermi surface properties in momentum space. For example, those parts of momentum space not associated with the VHs give a linear background dependence to the entropy evolution as a function of strain in the normal state. This contribution results in a small shift of the strain at which $(\partial s/\partial\varepsilon)_T=0$ and hence the entropy maximum relative to the strain at which the VHs crosses the Fermi energy.

Furthermore we would like to highlight the apparent kink in the evolution of the superconducting dome at a strain of approximately $-0.6\,\%$ that is seen in Fig.~\ref{fig:mappedfig6}. This is caused by the magnetic phase at higher strain coinciding with the superconducting phase transition at those strains~\cite{Grinenko2021,nature}. The question of wether superconductivity and magnetism coexist or compete has to be further investigated in future experiments extending to lower temperature.

\subsection{Entropy Analysis}

We next turn to the results for the total entropy shown in Fig.~\ref{fig:ST}. The data above the superconducting dome at high temperatures is qualitatively and quantitatively consistent with that reported by us previously in Refs.~\cite{nature,Palle2023}. For lower temperatures two kinks (broadened by strain inhomogeneity) appear as a function of strain as indicated by upward-pointing arrows in Fig.~\ref{fig:ST}. These represent the entrance into and exit from the superconducting phase. Note that the transition occurs at those points where $(\partial s/\partial\varepsilon )_T$ is changing fastest as a function of strain, i.e., where $(\partial^2 s/\partial\varepsilon^2 )_T$ is maximal. The entropy within the superconducting phase is lower than that expected from an extrapolation of the normal state evolution with strain. This is fully consistent with a superconducting gap in the electronic excitation spectrum. The further significance of the data lies in its absolute accuracy which will allow for quantitative comparison to theoretical modeling of the strain evolution of the superconducting order parameter candidates for \Sr ~\cite{Palle2023}.

\begin{figure*}[ht]
\centering
\includegraphics[width=\linewidth]{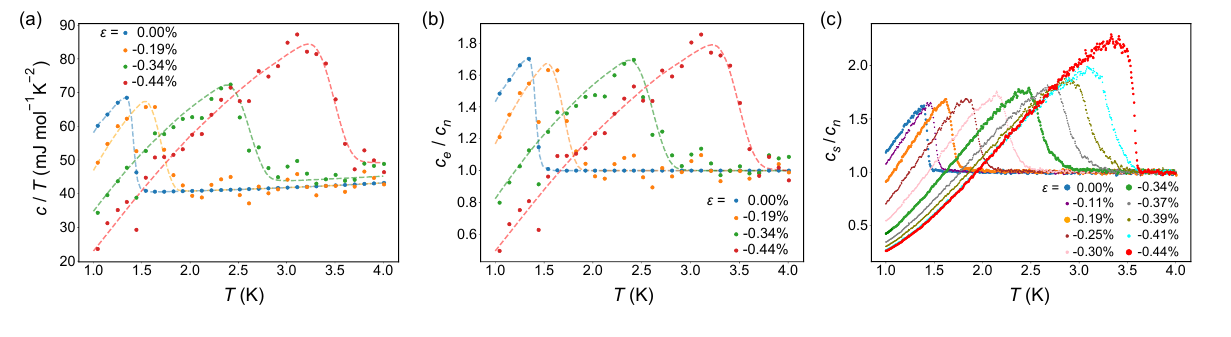}
\caption{Specific heat recovered from elastocaloric measurements.
(a) Specific heat divided by temperature, $c/T$, as a function of temperature at selected fixed strains.
In the reconstruction procedure, $\varepsilon = 0$ denotes the experimentally measured reference curve at zero applied strain.
(b) Normalized electronic specific heat, $c_e/c_n$, extracted at the same strain values with phonon contribution subtracted.
In panels (a) and (b), dashed lines are shown as guides to the eye.
(c) Normalized specific heat reproduced with data from Ref.~\cite{Li2021}.
The curves corresponding to the strain values shown in panels (a) and (b)
are highlighted with larger markers.}
\label{fig:HC}
\end{figure*}

A similar criterion based on the minimum of ($\partial^2 s / \partial \varepsilon^2)_T$ is used to define the onset of the magnetic state, as indicated by the solid arrows in the high compressive strain region ($\lvert \varepsilon \rvert >0.6\%$) of Fig.~\ref{fig:ST}. To estimate the reduction in entropy associated with the magnetic transition, we fitted the data at compressive strains below the magnetic transition onset with a quadratic function and extrapolated it to higher compressive strains, as shown by the dashed curves in Fig.~\ref{fig:ST}. This is the same procedure used in Ref.~\cite{nature} to estimate the entropy jump associated with the reduction in entropy connected with the magnetic transition above 4 K. Here, we extend this analysis to 2 K, below which the magnetic and superconducting transition signatures approach each other. We find no significant enhancement or reduction of the entropy jump divided by temperature upon cooling.

Finally, careful inspection of the entropy surface does not reveal any signatures that would be consistent with a further phase transition within the superconducting phase such as that indicated by $\mu$SR measurements~\cite{Grinenko2021}.
Even if the transition were effectively strain-independent, any entropy difference across it should result in a signature in the reconstructed entropy surface. As such, any remaining phase transition, if existing within the superconducting dome, would have an impact on the entropy of \Sr. This absence of any signature of a further phase transition is consistent with findings from the caloric measurements reported by Li \textit{et al.}~\cite{Li2021}.
This poses significant challenges to an interpretation of the $\mu$SR signal based on properties of the bulk superconducting phase as these generally would involve significant changes to thermodynamic properties.

\subsection{Specific Heat}

Finally, we turn to specific heat. Due to the high sensitivity and accuracy of the combined datasets it is possible to carry out a numerical differentiation with respect to temperature of the reconstructed entropy shown in Fig.~\ref{fig:ST}. We present the result of this calculation in Fig.~\ref{fig:HC}(a) where we show specific heat divided by temperature at constant strain $\left(c_\varepsilon/T\right)$ versus temperature (dots). In Fig.~\ref{fig:HC}(b) we present the same data, having subtracted the known phonon contribution~\cite{Mackenzie2003} and divided by the specific heat in the normal state just above the transition temperature for each strain. This presentation allows a more direct comparison to the data reported in \cite{Li2021} and replotted here in Fig.~\ref{fig:HC}(c). While the direct measurement in panel Fig.~\ref{fig:HC}(c) has superior signal-to-noise a number of approximations had to be made in order to analyze the raw data which were measured in the quasi-adiabatic a.c.\ specific heat regime~\cite{Li2021, Li2020}. As these result in systematic uncertainties on the absolute accuracy as well as the evolution over large temperature regimes Li \textit{et al.}\ focused predominantly on the evolution of the relative size of the jump in specific heat at the phase transition.

Our specific heat results based on the reconstructed entropy surface confirm some but not all of the key qualitative observations of Ref.~\cite{Li2021}, and add further information. First, our direct reconstruction of $c/T$ confirms that the normal state specific heat is increasing, in line with the increased entropy in this regime. We note that this is not a trivial statement on the magnitude of molar entropy $s$ but its temperature dependence and is a non-trivial result of the reconstruction of the entropy surface. No attempt was made to deduce this normal state heat capacity from the data presented in Ref.~\cite{Li2021}, because of the uncertainties discussed above. Secondly, the jump in $c/T$, $\Delta c/T$, at the superconducting phase transition has initially only a very weak dependence on strain once strain broadening is taken into account. Finally, close to $\varepsilon_\mathrm{VHs}$ both our and the more direct measurement show an increase in $\Delta c/T$.

However quantitatively the two measurements show significant differences in the increase in $\Delta c/T$ as determined from our elastocaloric effect based specific heat data is much smaller. A subsequent re-analysis of the caloric data published in Ref.~\cite{Li2021} revealed that it most likely overestimated the jump size at $\varepsilon_\mathrm{VHs}$ as it assumed that the thermal conductance of \Sr ~ does not vary significantly under strain. This assumption may be incorrect, as preliminary measurements of the thermal conductance reveal significant variations under strain. A comprehensive study of the thermal conductance will be discussed in a future publication.

\section{Conclusion}

In this paper, we present a quantitative framework for extracting absolute thermodynamic information from a.c.\ elastocaloric effect measurements under uniaxial pressure by leveraging information from both the quasi-adiabatic and strong coupling regimes, combining the high sensitivity of the former with the absolute accuracy of the latter.

We showed our ability to quantitatively reconstruct entropy changes both across phase transitions and as a function of strain and temperature across the full phase diagram. The high quality of the data thus obtained enables the determination of the entropy itself, as well as the subsequent temperature derivative, which yields specific heat. This allows us to draw significant conclusions about the superconducting properties of \Sr.

The elastocaloric response reveals that the strain that maximizes $T_\mathrm{c}$ does not coincide with the strain at which entropy is maximized. We attribute this behavior to Fermi surface anisotropy originating from the Van Hove singularity. Entropy reconstruction further confirms the absence of thermodynamically significant phase transitions within the superconducting dome. In addition, the specific heat analysis indicates a larger relative jump at the superconducting transition at the Van Hove singularity. However, that increase is much smaller than previously inferred from more direct heat capacity measurements which we believe require re-analysis. Finally, extending the entropy reconstruction down to lower temperatures enables us to track the onset of the magnetic phase into the regime where it approaches the superconducting dome. Within our experimental resolution, we observe no significant change in the temperature normalized entropy jump associated with the magnetic transition upon approaching the superconducting state.

In a more general context we demonstrate the capability of the elastocaloric effect in the strong coupling regime to be an absolute quantitative thermodynamic probe in uniaxial strain experiments where accurate specific-heat measurements are currently not possible. While we leveraged the thermodynamic properties of the low strain Fermi liquid regime for an absolute determination of the strained volume of the sample we also showed that this agrees to within 30~\% with a simple estimate based on optical measurements of the sample volume. The significance of this observation is that the latter approach is applicable to any material. Hence the presented method of exploiting a combination of strong-coupling and quasi-adiabatic measurements of the elastocaloric effect to reconstruct absolute thermodynamic properties will be a widely applicable route to quantitative mapping of entropy and specific heat across uniaxial pressure phase diagrams.

The data that support the findings of this study are openly available in the St Andrews Research Portal \cite{data}.

\begin{acknowledgments}
We are grateful to Y. Maeno for helpful discussions. We wish to acknowledge the Max Planck Society for financial support and the Deutsche Forschungsgemeinschaft (DFG, German Research Foundation) for funding through Grant No. TRR288--422213477, Project No. A10 (APM, HMLN). Research in Dresden benefits from the environment provided by the DFG Cluster of Excellence ctd.qmat (EXC 2147, Project ID 390858490). AWR acknowledges support from the Engineering and Physical Sciences Research Council (grant numbers EP/P024564/1 and EP/S005005/1). YSL acknowledges support from the National Science and Technology Council and the Ministry of Education under grants NSTC 114-2112-M-002 -019 -MY3, MOE-112-YSFMS-0003-003-P1 and NTU-CC-115L890402. NK acknowledges support from JSPS KAKENHI (No. JP21H01033, No. JP22K19093, and JP24K01461).
\end{acknowledgments}

\appendix

\section{Lumped-Element Model}
\label{app:model}
The second law of thermodynamics relates the entropy change $\mathrm{d}S$ of the whole sample to the
heat flow ${\rm \dbar} Q_s$ into or out of the sample according to
\begin{equation}
\label{master1}
{\rm \dbar} Q_s
= T_s {\rm d} S_\mathrm{tot},
\end{equation}
where $T_s$ is the sample temperature.

The entropy $S_\mathrm{tot}$ is a thermodynamic state function that depends only on strain
$\varepsilon$ and temperature $T_s$, i.e.\ $S_\mathrm{tot} = S_\mathrm{tot}(\varepsilon, T)$.
In the experiment, both $\varepsilon$ and $T$ vary with time, such that the
entropy acquires an implicit time dependence, $S_\mathrm{tot}(t) = S_\mathrm{tot}[\varepsilon(t), T(t)]$.

The heat flow into or out of the sample arises from two contributions:
the applied heating power from the heater ${\rm \dbar} Q_{\mathrm{heater}}$ and the thermal exchange with the
bath ${\rm \dbar} Q_{\mathrm{link}}$ through the thermal link.
Using Eq.~\ref{master1}, the entropy balance can therefore be written as
\begin{equation}
\label{master2}
\begin{split}
\frac{{\rm \dbar} Q_{\mathrm{heater}}}{{\rm d} t}
+ \frac{{\rm \dbar} Q_{\mathrm{link}}}{{\rm d} t}
={}&
T_s \left( \frac{\partial S_\mathrm{strain}}{\partial \varepsilon} \right)_{T}
\frac{{\rm d} \varepsilon}{{\rm d} t}
\\
&
+ T_s \left( \frac{\partial S_\mathrm{tot}}{\partial T_s} \right)_\varepsilon
\frac{{\rm d} T_s}{{\rm d} t}.
\end{split}
\end{equation}

\noindent Using the definition of the heat capacity at constant strain,
\begin{equation}
T_s\left(\frac{\partial S_\mathrm{tot}}{\partial T}\right)_\varepsilon
\equiv C_{\mathrm{tot}} = C_{\mathrm{strain}} + C_{\mathrm{unstrained}},
\end{equation}
and introducing the thermal conductance $k$ of the heat link to a bath held at
constant temperature $T_0$, Eq.~\ref{master2} can be rearranged as

\begin{equation}
\label{master3}
C_{\mathrm{tot}} \frac{{\rm d} T_s}{{\rm d} t}
=
\frac{{\rm \dbar} Q_{\mathrm{heater}}}{{\rm d} t}
- T_s \left( \frac{\partial S_\mathrm{strain}}{\partial \varepsilon} \right)_{T}\frac{{\rm d} \varepsilon}{{\rm d} t}
-
k (T_s - T_0).
\end{equation}
Here, the elastocaloric response is treated as an effective ``virtual'' heat source.
Explicitly, the corresponding elastocaloric power is given by
\begin{equation}
\label{power1}
\frac{{\rm \dbar} Q_{\mathrm{ECE}}}{{\rm d} t}
=
- T_s \left( \frac{\partial S_\mathrm{strain}}{\partial \varepsilon} \right)_T
\frac{{\rm d} \varepsilon}{{\rm d} t}.
\end{equation}

These expressions lead to the compact fundamental differential equation
\begin{equation}
\label{ECEmaster}
C_{\mathrm{tot}} \frac{{\rm d} T_s}{{\rm d} t}
=
\dot{Q}_{\mathrm{ECE/heater}}
-
k (T_s - T_0),
\end{equation}
which describes both heating- and elastocaloric effect-driven measurements within a
unified framework, making their formal equivalence explicit.

The above differential equation can be solved analytically and in the case of the elastocaloric mode results in temperature oscillations $\tilde{T}_s \cos(\omega t)=T_s-T_0$ as a function of frequency $\omega$ given by
\begin{equation}
\left|\tilde{T}_s\right|=
\frac{\omega}{\sqrt{1+\omega^2\left(C_\mathrm{tot}/k\right)^2}}\times\frac{T_0}{k}\left|\left(\frac{\partial S_\mathrm{strain}}{\partial \varepsilon}\right)_T\tilde{\varepsilon}\right|,
\end{equation}

\noindent which has a characteristic time / frequency scale set by $\tau_{\rm 1}=1/\omega_{\rm 1}=\frac{C_\mathrm{tot}}{k}$. The functional form is shown in the blue solid curve in Fig.~\ref{fig:fig3}.

\bibliography{Entropy.bib}

\begin{thebibliography}{27}
\bibitem [{ent()}]{entropy_compliance}%
  \BibitemOpen
  \href@noop {} {}\bibinfo {note} {Please note that we elected to keep the
  standard notation of $s$ for molar entropy, $\mathbf{s}$ for the compliance
  matrix as the former only appears as a scalar and the latter as a matrix
  throughout the paper.}\BibitemShut {Stop}%
\end{thebibliography}

\end{document}